\documentclass[twocolumn,superscriptaddress,amsmath,amssymb,nofootinbib,nobibnotes,physrev,floatfix]{revtex4-2}
\usepackage[utf8]{inputenc}
\usepackage[T1]{fontenc}
\usepackage[version=3]{mhchem}
\usepackage{graphicx}  
\usepackage{dcolumn}   
\usepackage{bm}        
\usepackage{units}
\usepackage[english]{babel}
\usepackage[dvipsnames]{xcolor}
\usepackage[
colorlinks=true,
urlcolor=RoyalBlue,
linkcolor=RoyalBlue,
citecolor=RoyalBlue,
]{hyperref}
\usepackage{vmargin}
\setpapersize{A4} \oddsidemargin20mm \textwidth170mm \topmargin10mm
\usepackage[type1]{libertine}
\usepackage{textcomp}
\usepackage[scaled=.85]{beramono}
\usepackage[libertine,cmintegrals,cmbraces,vvarbb]{newtxmath} 
\usepackage[scr=boondoxo]{mathalfa}

\usepackage[lf]{carlito}
\newcommand\blfootnote[1]{%
  \begingroup
  \renewcommand\thefootnote{}\footnote{#1}%
  \addtocounter{footnote}{-1}%
  \endgroup
}

\begin{document}
\title{Character of electronic states in the transport gap of molecules on surfaces}
\author{Abhishek Grewal}
\blfootnote{This document is the Accepted Manuscript version of a Published Work that appeared in final form in ACS Nano, after peer review and technical editing by the publisher. To access the final edited and published work see \url{https://doi.org/10.1021/acsnano.2c12447}, ACS Nano, {\bf 17}, 14, 13176–13184 (2023).\\}
\author{Christopher C. Leon}

\author{Klaus Kuhnke}
\email{k.kuhnke@fkf.mpg.de}
\affiliation{Max-Planck-Institut f{\"u}r Festk{\"o}rperforschung, Stuttgart, Germany}

\author{Klaus Kern}
\affiliation{Max-Planck-Institut f{\"u}r Festk{\"o}rperforschung, Stuttgart, Germany}
\affiliation{Institut de Physique, {\'E}cole Polytechnique F{\'e}d{\'e}rale de Lausanne, Lausanne, Switzerland}

\author{Olle Gunnarsson}
\email{o.gunnarsson@fkf.mpg.de}
\affiliation{Max-Planck-Institut f{\"u}r Festk{\"o}rperforschung, Stuttgart, Germany}

\begin{abstract}
We report on scanning tunneling microscopy (STM) topographs of individual metal phthalocyanines (MPc) on a thin salt (NaCl) film on a gold substrate, at tunneling energies within the molecule's electronic transport gap.
Theoretical models of increasing complexity are discussed.
The calculations for MPcs adsorbed on a thin NaCl layer on Au(111) demonstrate that the STM pattern rotates with the molecule's orientations $-$ in excellent agreement with the experimental data.
Thus, even the STM topography obtained for energies in the transport gap represent the structure of a one atom thick molecule.
It is shown that the electronic states inside the transport gap can be rather accurately approximated by linear combinations of bound molecular orbitals (MOs).
The gap states include not only the frontier orbitals but also surprisingly large contributions from energetically much lower MOs.
These results will be essential for understanding processes, such as exciton creation, which can be induced by electrons tunneling through the transport gap of a molecule.
\end{abstract}

\maketitle
\section*{Introduction}
Imaging molecules on surfaces with scanning tunneling microscopy (STM) often involves resonant tunneling through its electronic molecular orbitals (MOs).
This process leads to an extremely enhanced tunneling rate which facilitates high-resolution imaging of specifically chosen electronic orbitals.
This mechanism is experimentally and theoretically well established
\cite{chenIntroductionScanning2021, reppMoleculesInsulating2005, reppControllingCharge2004c, liljerothSingleMoleculeSynthesis2010}.
In contrast, off-resonant tunneling through the transport gap between two MOs can show interesting behaviors that venture far beyond this standard.
This situation becomes particularly important when the two MOs are the highest occupied MO (HOMO) and the lowest unoccupied MO (LUMO), and tunneling through the energy gap is used, for example to create singlet excitons for photon emission \cite{qiuVibrationallyResolved2003, kuhnkeAtomicScaleImaging2017, miwaManyBodyState2019, chenSpinTripletMediatedUpConversion2019a, farrukhBiaspolarityDependent2021, grewalSingleMolecule2022}.
Tunneling through a transport gap occurs also for devices with negative differential resistivity \cite{tuControllingSingleMolecule2008}.

The importance of these fundamental processes leads us to examine the details of electron tunneling within the molecule's electronic transport gap.
We focus on the electron propagation from the substrate to the molecule.
The molecule studied here, platinum(II) and magnesium phthalocyanine (PtPc and MgPc), are one atom thick molecules, the thinnest possible.
Nevertheless, we find experimentally that the STM topography image is decisively influenced by the molecule, even for tunneling at energies in the electronic transport gap where the molecule is non-conducting.
However, the images in the entire transport gap differ strongly from the images of the HOMO and LUMO, even when these orbitals are just a few tens of meV away from the tunneling electron energy.
We find, theoretically, that the gap images of the molecule can be described to a good approximation by linear combinations of bound MOs.
Surprisingly, we find that MOs at energies far below the gap play an essential role in the gap images, explaining why they look substantially different from both the HOMO, and the LUMO.

In STM or STM-induced luminescence studies, the molecule is often deposited on a few layers of a large band gap insulator, such as NaCl \cite{reppMoleculesInsulating2005, qiuVibrationallyResolved2003, cavarFluorescencePhosphorescence2005}.
The insulator is often considered as an uninteresting buffer, simply present to make the coupling between the molecule and the substrate weak, but is otherwise not very important.
In a recent paper, we have shown that the conduction band of NaCl has mainly Cl character, like the valence band, contrary to common assumptions \cite{leonAnionicCharacter2022, grewalSingleMolecule2022}.
The gap electrons then also have wave functions of mainly Cl character in the NaCl film, which influences the coupling of a molecule to NaCl.
This important aspect is taken into account in our calculations.
The coupling between the Au(111) substrate and the PtPc molecule via the NaCl film strongly favors specific PtPc MOs, which play an important role in the topography imaging at energies in the electronic transport gap of PtPc.

We perform a set of calculations for models of increasing complexity.
The purpose is to explain why gap images are strongly influenced by the MOs, even at energies in the transport gap. In particular, we consider an exactly solvable model of a substrate and an adsorbed molecule with a HOMO and a LUMO. We show that in the spatial range of the molecule, the wave function to a good approximation is a linear combination of the HOMO and LUMO, even for tunneling through the transport gap. This does not imply any violation of energy conservation whatsoever since the HOMO and LUMO are not eigenfunctions of the combined system - molecule with substrate. The calculations illustrate that the MOs provide a very good basis set.

We then perform realistic model calculations for PtPc and MgPc adsorbed atop three layers of  NaCl(100) on an Au(111) substrate, using all the MOs as an efficient basis set for expanding the wave function inside the molecule.
The experimental topographic images of electrons tunneling through the gap are reproduced rather accurately.

This effort revealed that absolute rotational orientation, adsorption site, and metal center are important, in this order, to the gap images of these molecules.
It is specifically dominated by orientation, spotlighting the importance of the MOs even for tunneling through the gap.

\section*{Results and Discussion}
\subsection{Theoretical discussion of electron propagation through the transport gap}
\begin{figure}
\includegraphics[width=\linewidth]{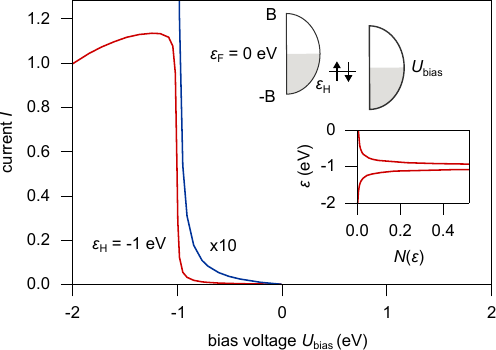}
\caption{Tunneling current for a model consisting of a substrate, a tip and a molecule with a HOMO at energy $\varepsilon_\text{H}=-1$ eV.
The upper insert shows the model and the lower insert shows how the coupling between the HOMO and the substrate leads to a narrow resonance in the local density of states, $N(\varepsilon)$, on the HOMO.
The hopping matrix elements between the molecule and the tip are turned on at large negative times at the rate $e^{\kappa t}$.
The current (red curve) at time $t=0$ is shown as a function of $U_\text{bias}$, with $\kappa\hbar=0.01$ eV, where $\hbar$ is the reduced Planck constant.
The current is normalized to unity at $U_\text{bias}=-2$ V, and its value multiplied by ten is shown by the blue curve.
The figure illustrates that there is a small current to the tip even for $\varepsilon_\text{H}<U_\text{bias}<0$, due to the broadening of the HOMO. We have used the bandwidth $2B=4$ eV for both the substrate and the tip.
The temperature is $T=0$, and the tail at higher energy is therefore not due to thermal effects.}
\label{fig:1}
\end{figure}

To improve our understanding of gap states, we first consider a straightforward tight-binding model.
As shown in the inset of Figure \ref{fig:1}, we consider a molecule with just one orbital (HOMO), at the energy $\varepsilon_\text{H}<0$ eV, on a substrate, and its coupling to a metal tip.
The voltage bias is $U_\text{bias} \le 0$ eV.
We include hopping matrix elements from each substrate level to the HOMO level and from the HOMO level to each tip level.
We first calculate the states of the system without the tip.
$N(\varepsilon)$ shows the corresponding local density of states (DOS) on the HOMO.
There is a narrow resonance around $\varepsilon_\text{H}$, but with tails extending to energies far away from $\varepsilon_\text{H}$.
The hopping integrals between the molecule and the tip are turned on at some large negative time with a slow growth, $e^{\kappa t}$, where we let $\kappa \rightarrow 0^+$, to some very small positive value.
The computed results are shown in Figure \ref{fig:1}.
For further details, such as the effects of introducing the Coulomb interaction, and how the hopping between the molecule and the tip is treated in first-order perturbation theory, see the Supporting Information (SI).

For $U_\text{bias}<\varepsilon_\text{H}$, there is a large current, as expected, since the tip Fermi energy is below the unperturbed HOMO level.
The drop in current as the bias is made more negative reflects the semi-elliptic form of the DOS.
However, even for $\varepsilon_\text{H}<U_\text{bias}<0$, there is a non-vanishing current due to a small Lorentzian tail of the narrow resonance for $\varepsilon>\varepsilon_\text{H}$.
Away from the resonance, the tail decays rather slowly as $1/(\varepsilon-\varepsilon_\text{H} )^2$.
We emphasize that this current, however small, is not negligible.

Tunneling through the HOMO for $\varepsilon_\text{H} < U_\text{bias}$ does not imply violation of energy conservation, since the HOMO is not an eigenstate of the Hamiltonian describing the combined substrate-HOMO system.
In the following we show that this tunneling through the HOMO and, in particular, tunneling through lower-lying MOs as well as the LUMO is crucial for understanding the image of electrons tunneling through the molecule's transport gap.
This set of considerations then provides a unified and consistent description of tunneling for all values of the bias voltage.

We now discuss two essential assumptions in the model above.
Firstly, the current flows entirely via the HOMO even for $\varepsilon_\text{H}<U_\text{bias}$, since there is no direct hopping from the substrate to the tip.
The tip then sees the lateral structure of the HOMO of the molecule, and it does not see the structure of the substrate.
This is true even when $\varepsilon_\text{H}<U_\text{bias}$ and the resulting hole has almost all the weight in the substrate.
Secondly, we have assumed that there is only one orbital on the molecule.
Including several orbitals would allow for interesting interferences effects between the hopping through different MOs.

\begin{figure}
\includegraphics[width=\linewidth]{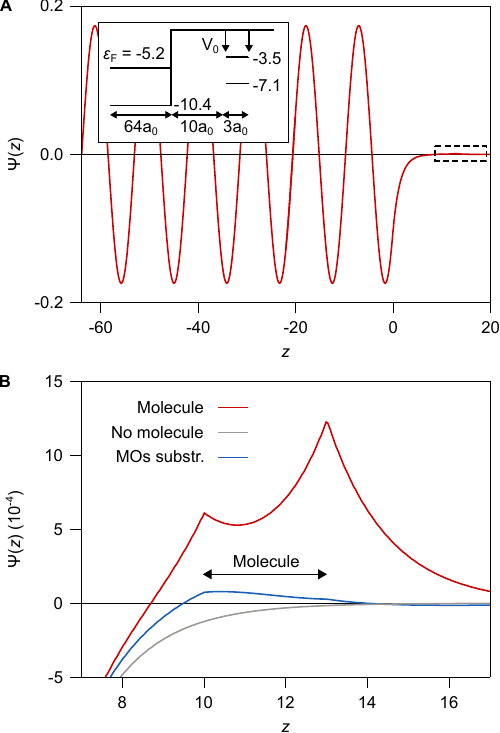}
\caption{(A) Wave function for an energy ($-6.3$ eV) in the energy gap of the molecule as a function of spatial coordinate $z$, which is in units of $a_0$.
The substrate is at the left $(z<0)$ and a molecule with nuclei at $z=10$ and $z=13$ to the right (see insert).
The HOMO and the LUMO are located at $-7.1$ eV and $-3.5$ eV, respectively.
The amplitude of the wave function on the molecule is very small and barely visible on this scale.
(B) Blow up of A in the region of the molecule.
The figure shows how the amplitude of the wave function is hugely enhanced (factor 78 at $z=17$) when the molecule is included (red curve) compared with the case without a molecule (gray curve).
It also shows the small fraction of the exact wave function which cannot be expanded in the HOMO and LUMO wave functions in the range $(10\le z \le13)$ of the molecule (blue curve).}
\label{fig:2}
\end{figure}

To discuss these assumptions, we study a one-dimensional $(1d)$ model which can be solved exactly, so that there is no need to introduce a basis set or make assumptions about hopping matrix elements.
This model is shown schematically in the inset of Figure \ref{fig:2}A.
To the left is a substrate $(-64 \le z \le 0)$ with the surface at $z=0$, and to the right is a simplified molecule with two nuclei at $z=10$ and $z=13$, with the spatial coordinate $z$ in Bohr radii $(a_0)$.
The substrate has the Fermi energy at $-5.2$ eV and a potential of $-10.4$ eV.
The nuclei of the molecule are described by two $\delta$-functions whose intensities set the HOMO and LUMO at $-7.1$ eV and $-3.5$ eV, respectively.
The specific energies here were chosen to represent PtPc adsorbed on Au.

Figure \ref{fig:2}A shows a wave function solution for one energy ($E^*=-6.3$ eV) in the gap.
Figure \ref{fig:2}B shows the same solution in detail, in the range of the molecule.
The wave function in the presence of a molecule (red curve) is hugely enhanced (by a factor of 78 at $z=17$) compared with the solution (gray curve) without a molecule.

It is sometimes heuristically suggested that in the energy range of the gap of the molecule, owing to the absence of MOs, the solution would be strongly reduced by the presence of the molecule.
To see that in general this cannot be the case, we discuss the difference between a free molecule and an adsorbed molecule.
For a free molecule, we require that the solution decays exponentially on both sides of the molecule for $\varepsilon<0$.
Considering this free molecule, one realizes that this exponential decay can only be satisfied at exactly two energies, corresponding to the HOMO and the LUMO.
For all other energies, $\varepsilon<0$, the wave function grows exponentially unbounded on at least one side of the molecule, and it is, therefore, not a physically admissible solution.
When the molecule sits in the presence of a solid, however, the wave function is allowed to be (and typically is) exponentially growing on the side facing the solid (and exponentially decaying as seen from the perspective of the solid), and therefore, energies in the gap are allowed.
The presence of the solid completely changes the character of admissible wavefunctions, no matter how “weakly” it may perturb the system.
Indeed, typically the presence of the molecule hugely enhances the wave function amplitudes rather than suppresses it.
In fact, for the illustrative energy $E^*$ in Figure \ref{fig:2}B, its associated wave function even grows with $z$ inside the molecule.
When this wave function is compared with those associated with energies close to the HOMO or the LUMO, it is of course strongly reduced, as is seen in Figure \ref{fig:1}.
Although an example is not illustrated in Figure \ref{fig:2}, it is possible, however, to choose the orbital energies such that the tunneling is reduced by presence of the molecule for some energies.

Similar arguments apply to the NaCl film.
For an infinite NaCl solid there are no physical states in the band gap.
For the present system, however, Au states have exponentially decaying tails extending through the NaCl film and the molecule out to the tip, even for energies corresponding to the gaps of NaCl and PtPc.

In the context of this model, the first assumption above implies that we only need to consider the indirect coupling between the substrate and the tip via the HOMO and LUMO.
The blue curve in Figure \ref{fig:2}B shows
\begin{equation}
\Psi(z)-\sum_{i=1}^2 \Phi_i(z) \int \Phi_i(z^\prime) \Psi(z^\prime). dz^\prime
\end{equation}
The second term represents the expansion of the exact wave function using the bound solutions $\Phi_i (z)$ of the free molecule.
The blue curve in Figure \ref{fig:2}B illustrates that in the range of the molecule, just a small remainder of $\Psi(z)$ cannot be expanded in the bound solutions of the free molecule.
Comparing the gray and red curves, we observe that the presence of the molecule and its attractive potential hugely enhances the wave function amplitude which results in an increased probability to find the electron at the position of the molecule.
Given this result, that the molecule's presence hugely enhances the wave function amplitude, it is not surprising that the exact wave function primarily consists of a linear combination of the HOMO and LUMO in the range of $z$ that overlaps with the molecule.

This picture then justifies the first assumption that there is no direct hopping from the substrate to the tip; the tip just couples to the states on the molecule.
In the example above, this means neglecting the coupling of the tip to the small residual (blue curve) in Figure \ref{fig:2}B.
In the model calculation to follow, we assume that this neglect remains a good approximation.
In the full three-dimensional case, orbitals on the tip with a specific symmetry may additionally dominate the hopping.
It is then essential how the important orbitals of the molecule and the underlying substrate couple to the tip orbitals, as such couplings can also strongly influence the tunneling from the molecule.
Finally, we notice that in the three-dimensional case, there can also be direct tunneling of electrons from the Au/NaCl system to the tip without passing through the molecule.
This contribution is neglected here.
Further details of the model are presented in the SI.

Concerning the second assumption above, that there is only one orbital on the molecule, we observe that in the range of the molecule, the wave function is now a linear combination of two functions.
In the SI, we show that this results in a strong energy dependence of the wave function, which can easily be understood in terms of the coupling to the two MOs, in strong contrast to the simple model in Figure \ref{fig:2}, which only has one MO.
Although the direct coupling between the substrate and the tip may be minimal, it is indirectly affected by coupling via different MOs.
For the PtPc model studied below, 182 states on the molecule lead to a rich coupling to the substrate.

We can now perform a much more realistic calculation for molecules on a gold substrate covered by a three-layer NaCl film.
We study PtPc experimentally and theoretically and compare theoretical results for MgPc with experimental results by Miwa \textit{et al.}~\cite{miwaEffectsMoleculeinsulator2016a}
We use a tight-binding model for Au, including $3d$, $4s$, and $4p$ orbitals. 
For the NaCl film, we include the Na $3s$ and $3p$ orbitals.
As discussed in our earlier work \cite{leonAnionicCharacter2022}, the conduction band of NaCl is primarily of Cl character, in contrast to common belief that the conduction band is cationic in character.
We thus include the Cl $3p$ and $4s$ levels and adjust the parameters so that the conduction band is mainly of Cl $4s$ character.
The model for the Au-NaCl system is identical to the model in Ref. 13. We then add a model of the adsorbed molecule, not included in the earlier work.
The PtPc or MgPc molecule is described by including all 57 atoms.
We use the empirical parameters of Harrison \cite{harrisonElementaryElectronic1999}, but we have modified the parameters slightly, for e.g., to obtain the experimental PtPc HOMO-LUMO energy gap, including image effects, and to obtain the correct alignment of electronic structures in the sub systems.
We did not tune parameters in order to improve the agreement with the experimental images.
For details of the parameters employed, see the SI.

\begin{figure}
\includegraphics[width=\linewidth]{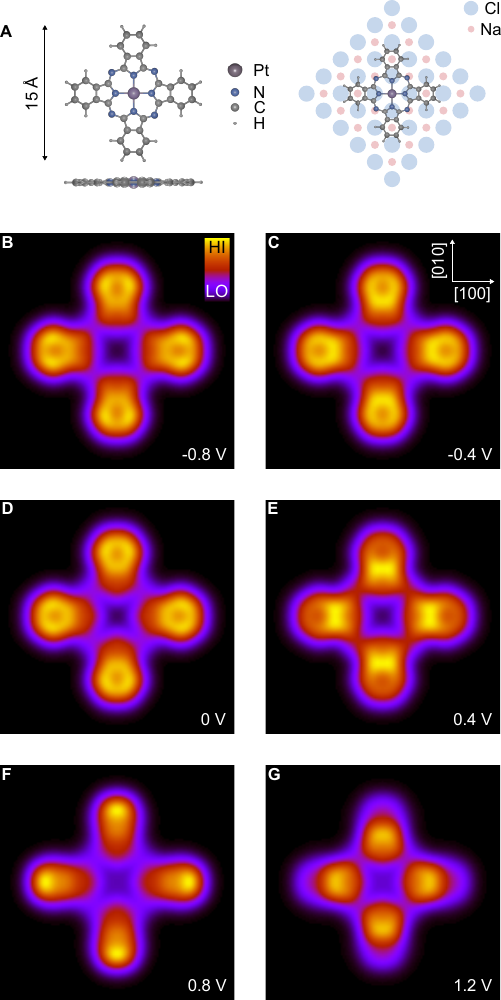}
\caption{(A) Left panel: Ball and stick model of the PtPc molecule in top view and side view.
Right panel: Top view of the adsorption geometry of PtPc on the NaCl layer.
(B-G) Theoretical PtPc images at the energies indicated in the lower right corner of each panel.
The images are calculated 1 \AA~ outside the molecular plane of PtPc adsorbed on a three layer NaCl(100) film on Au(111).
The crystallographic axes of NaCl are indicated in panel C. Images sizes $(16\times16$ \AA$^2)$.}
\label{fig:3}
\end{figure}

The corresponding one-particle Hamiltonian is solved for energies in the gap of PtPc or MgPc.
Even for tunneling through the transport gap, this approach allows for charge fluctuations on the molecule.
To obtain STM images, we use the Slater \cite{slaterAtomicShielding1930} rules to construct orbitals on the atoms, which are combined with the eigenvectors of the Hamiltonian.
For the interesting energy range, most of these wave functions are $\pi$-orbitals, i.e., mainly linear combinations of C and N $2p_z$ orbitals. For distances close to the molecular plane, these functions should provide a reasonable basis set.
In what follows, we will focus on images at these distances, but also show images for a realistic tip-sample distance of 7 \AA~as determined by point contact measurements \cite{jaculbiaSinglemoleculeResonance2020}.
For this purpose we introduce the approximations of Tersoff and Hamann \cite{tersoffTheoryApplication1983, tersoffTheoryScanning1985}, making it sufficient to calculate the electron wave function at a fictitious center of an $s$-orbital on the tip.
We assume that the potential in vacuum is constant inside a cylinder with radius 12 \AA~and infinite outside.
This radius is much larger than the distance from the cylinder axis to the outermost H atoms (7.6 \AA).
It is then a good assumption to assume that the wave function of the tunneling electrons is localized within the cylinder.
The Schr{\"o}dinger equation in vacuum is solved using a basis set.
We use functions $e^{\pm m\phi}$ to describe the angular dependence, where $m$ is an integer.
The radial behavior is described by integer Bessel functions and the behavior perpendicular to the surface by exponential functions, $e^{-\kappa z}$, where $\kappa$ is related to the energy of the electron.

The contact of the PtPc molecule to the rest of the system means that the PtPc charge  
is not conserved.
As a result, PtPc has charge fluctuations.
Projecting out the NaCl states in perturbation theory and considering states within $\pm 3$ eV of the Fermi energy, we obtain fluctuations out of neutral PtPc of the order of $10^{-3}$.

\subsection{Comparison between theory and STM measurements}
\begin{figure}
\includegraphics[width=\linewidth]{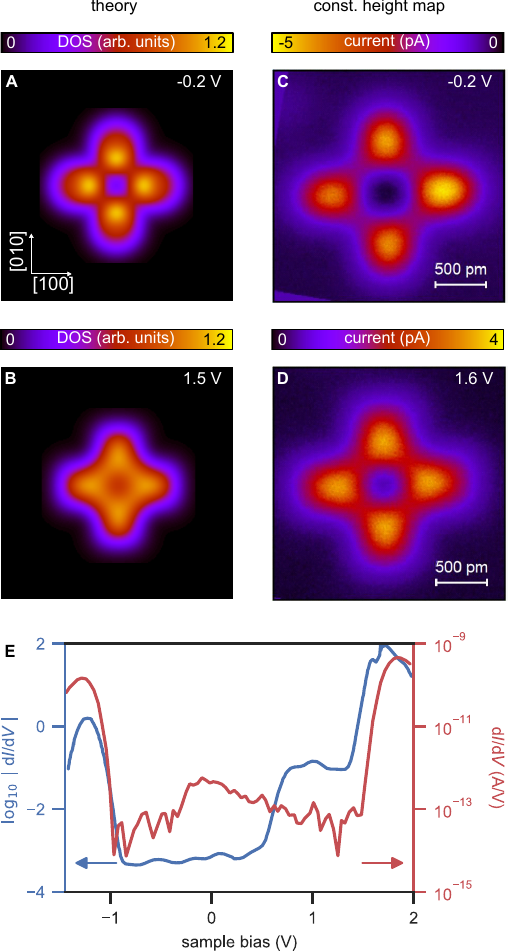}
\caption{Comparison of theoretical (A, B) and experimental (C, D) images for PtPc atop three layers of NaCl on Au(111) at energies given in the upper right of each panel.
The tip-molecule distance in the calculation is 7 \AA.
The experimental images are constant height STM maps.
The length scale of all panels is given in the panels C and D (image sizes $23\times23$ \AA$^2$).
Linear color scales are used for both experimental and theoretical data.
The difference in apparent molecular size is ascribed to the finite tip radius in experiment, which is not accounted for in the calculation.
(E) Logarithm of the calculated differential conductance $\left|{\rm d}I/{\rm d}V\right|$ as a function of bias $V$ at a tip-molecule distance 7 \AA~ (solid black line) in comparison to an experimental d$I$/d$V$ spectrum (blue markers).
For details see the SI.}
\label{fig:4}
\end{figure}

Calculations are performed for models of PtPc (Figure \ref{fig:3}A) and MgPc on a trilayer NaCl(100) film on Au(111).
For details of the parameters used, see the SI.
Computed results for PtPc at a distance of 1 \AA~are shown in Figure \ref{fig:3}B-G for different values of the energy $\varepsilon$ inside the gap.
The theoretical images exhibit four lobes on the isoindole units of the molecule, similar to experimental observation and in stark contrast to maps of both HOMO and the two overlapping degenerate LUMOs, which have eight lobes \cite{grewalSingleMolecule2022, wangReviewArticle2012}.
In Figure \ref{fig:3} the small changes in the theoretical results as a function of energy may be due to details of the calculation.
They are not found in experiment, even when the tip-molecule distance is as small as stable scanning permits.
Figure \ref{fig:4}A, B shows calculations at a more realistic tip-sample distance \cite{jaculbiaSinglemoleculeResonance2020} of $z=7$ \AA~ in comparison with constant height STM maps exhibiting a satisfactory agreement with experiment.
The energies studied in Figure \ref{fig:4}B, D are close to the LUMO ($\varepsilon=1.7$ eV) (see the density of state spectra in Figure \ref{fig:4}E) and demonstrate the amazingly rapid change from the orbital patterns to the gap images.
The difference in size of the computed images and the STM topography are ascribed to the limited resolution of the experiment, which arises from the finite tip curvature.

\begin{figure}
\includegraphics[width=\linewidth]{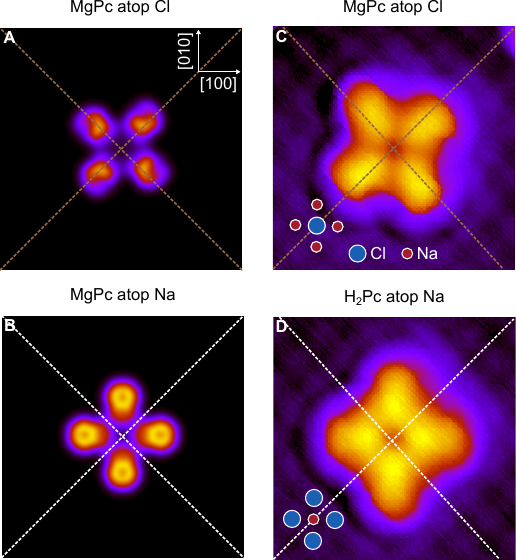}
\caption{Comparison of theoretical (A, B) and experimental (C, D) images for MgPc and \ce{H_2Pc} on bilayer NaCl on Ag(111) at $-0.55$ eV energy ($32\times32$ \AA$^2$).
Panels (C, D) show constant current ($I$ = 1 pA) topographic STM scans using raw data from a study
by Miwa \textit{et al.} \cite{miwaEffectsMoleculeinsulator2016a}
The MgPc molecule in (C) is adsorbed on a Cl site and is rotated by $53^\circ$ with respect to the NaCl (010) axis.
Both effects are included in the respective calculation (A).
\ce{H_2Pc} (as does PtPc) adsorbs on a Na site and is aligned with the NaCl lattice (D).
This situation was included in the calculation for a tip distance of 7 \AA~ above the molecule by orienting MgPc to the correct \ce{H_2Pc} adsorption geometry.
The apparent molecule sizes of experiment and theory are closer to each other than in Figure \ref{fig:4} since the experiment by Miwa \textit{et al.} used a CO-covered tip, known to provide sharper imaging.
Finite tip size and the effect of CO are not accounted for in the calculation presented in the current work.}
\label{fig:5}
\end{figure}

Figure \ref{fig:5} shows theoretical (Figures \ref{fig:5}A and B) and experimental \cite{miwaEffectsMoleculeinsulator2016a} (Figures \ref{fig:5}C and D) results for \ce{MgPc} and \ce{H_2Pc}.
The experimental data is obtained using a carbon monoxide molecule decorated tip which is known to improve STM spatial resolution \cite{hapalaOriginHighResolution2014}.
MgPc differs from PtPc in three ways.
An Mg atom has replaced the central Pt atom, the molecule is adsorbed on a Cl atom site of NaCl instead of a Na site, and the molecular orientations on the NaCl differ substantially.
As shown by Miwa \textit{et al.} \cite{miwaEffectsMoleculeinsulator2016a}, MgPc is oriented approximately $53^\circ$ off the (010) axis of the underlying NaCl lattice (Figure \ref{fig:5}C) in contrast to PtPc which is aligned with this axis.
\ce{H_2Pc} on the other hand has no central metal atom and adsorbs atop a Na atom, similar to PtPc.
All images in Figures \ref{fig:3}, \ref{fig:4} and \ref{fig:5} are oriented such that the horizontal and vertical directions of the image correspond to the NaCl (010) and (100) axes.

The theoretical images of MgPc (Figures \ref{fig:5}A and B) well reproduce the four lobe structure with a central minimum experimentally observed by Miwa \textit{et al.} \cite{miwaEffectsMoleculeinsulator2016a} (Figure \ref{fig:5}C).
In particular, the image in Figure \ref{fig:5}A is rotated relative to the PtPc image.

To check if the essential difference between MgPc and \ce{PtPc}/\ce{H2Pc} lies in the differences in molecular adsorption geometry, MgPc is purposely rotated in the calculation (Figure \ref{fig:5}B) so that its orientation is the same as observed for \ce{PtPc} and \ce{H_2Pc}, that is, along the (010) axis of NaCl.
Finally, the MgPc center is placed atop a Na atom.
The result is shown in Figure \ref{fig:5}B.
The image is now rather similar to the image for PtPc. We conclude that the difference in the molecule's orientation is the most crucial difference between PtPc and MgPc.
It is striking that the orientation of the molecule is so crucial, even for tunneling in the conduction gap.

\begin{figure}[b]
\includegraphics[width=\linewidth]{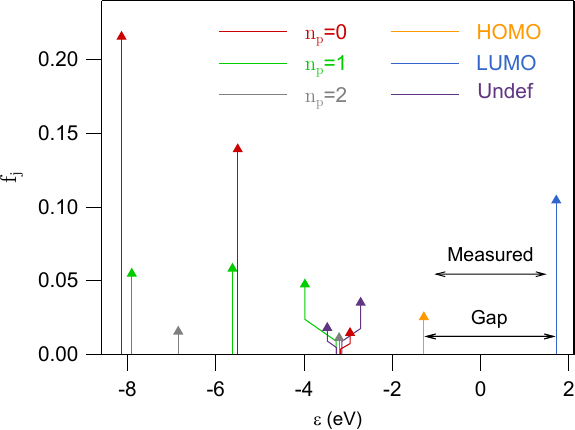}
\caption{Weights $f_j$ (Eq. 3) of important MOs in PtPc summed over all states within the gap between $\varepsilon=-0.8$ eV and 1.2 eV.
The energy range over which the sum extends, is marked by the arrow “Measured”.
Some $\pi$ states are labeled by the number of nodal planes, $n_p$.
States with more complicated patterns are labeled “Undef”.
The states represented in this figure contribute 74\% of the total weight.
The remaining contribution comes from many MOs each with rather small weights.
The $n_p$=1 states are doubly degenerate, and the weights of the degenerate states were added.}
\label{fig:6}
\end{figure}

To understand the shape of the images in Figures \ref{fig:3} and \ref{fig:4}, we expand the wave function of the combined system in terms of the PtPc MOs.
Inside the molecule, we write
\begin{equation}
	|\Psi_i\rangle = \sum_{j=1}^{182} c_j^{(i)}|j\rangle,
\end{equation}
where the sum is over the 182 eigenfunctions of the free molecule.
We then focus on ``diagonal'' contributions
\begin{equation}
	f_j = C\sum_{-0.8\le\varepsilon_i\le1.2}|c_j^{(i)}|^2,
\end{equation}
where $C$ is chosen such that $\sum_j f_j=1$.
Here $f_j$ adds up the weight on the PtPc's $j$th MO over all states within the energy range $-0.8$ eV $\le\varepsilon_i \le 1.2$ eV.
We observe, however, that “off-diagonal” contributions $\left[c^{(i)}\right]_j^* c_j^{(i^\prime)}$, $i \neq i^\prime$, also give substantial contributions.
Figure \ref{fig:6} shows $f_j$ for important MOs of $\pi$-character.
The $\pi$-states are labeled by the number of angular nodal planes $n_p$, i.e., planes through the center of the molecule and perpendicular to molecule and surface plane.
In cases where such nodal planes are not well defined, we have labeled the corresponding state ``Undef''. States with a given value of $n_p$ have different numbers of ``radial'' nodes assuring orthogonality.
Although the margins of the energy range approach the HOMO (at -1.3 eV) on one side and the LUMO (at 1.7 eV) on the other side to within 0.5 eV, HOMO and LUMO contribute only 3\% and 10\%, respectively, to the total weight.
Next, we have selectively summed up only contributions from states with a well-defined $n_p$-value.
The results are shown in Table 1.
Interestingly, three $n_p=0$ states (37\%) and six (including degeneracy) $n_p=1$ states (16\%) contribute almost half of the weight (53\%).
The $n_p=2$ states contribute little (3\%).
States with less well-defined angular nodes, shown in Figure \ref{fig:6}, contribute 5\%.
Many other states, have smaller contributions and are not shown in the figure.
Together they account for 26\%.

The two-fold degenerate $n_p=1$ states have leading contributions of the type $\sin^2(m\phi)$ and $\cos^2(m\phi)$ with $m=n_p=1$, where $\phi$ is the azimuthal angle.
They are planar two-lobe structures that lie along the $y$- and $x$-axis, respectively.
When combined they provide an approximately $\phi$-independent, isotropic contribution, just like the $n_p=0$ states.
The weak four-fold pattern is partly due to a $n_p=2$ function, with the symmetry $(x^2-y^2)^2$ and 4 lobes directed along the cardinal directions.
However, there are also contributions to the image from products of functions with different values of $n_p$, e.g., of the type $\cos(n_p\phi) \cos(n_p^\prime\phi)$, where $n_p=0$ and $n_p^\prime=4$.
Such functions are positive for multiples of $90^\circ$ and thus add weight along the $x$- and $y$-axis but subtract weight along the diagonals.
These images in the energy gap are very different from, e.g., the HOMO $(n_p=4)$, which is described by $m=4\nu$ $(\nu=1, 2, \dots)$ states, and the LUMO, which is described by odd $m$-value states with a significant weight for $m=3$ and $m=5$ (for illustration see refs. \onlinecite{grewalSingleMolecule2022} and \onlinecite{wangReviewArticle2012}).

Figure \ref{fig:6} illustrates that the Au-PtPc coupling via NaCl is far from trivial.
NaCl provides a buffer between the Au substrate and the PtPc molecule, but it influences the coupling in non-uniform ways, favoring the coupling to specific MOs.
This has implications for STM topography imaging in the PtPc transport gap.

\begin{table}
\caption{Relative contributions (weights) to the gap states in the interval $-0.8\le\varepsilon\le1.2$ eV.
Listed are the weights of some $\pi$-orbitals with different $n_p$-values, and the HOMO and LUMO orbitals.
“Other” shows the contributions in Figure \ref{fig:6} from orbitals that were not assigned an $n_p$ value, and “Rest” shows the many small contributions not shown in the figure.
The weights represent the total weights of a given state, not just from the leading $m$-component.\\}
\label{tbl:1}
\begin{tabular}{ccccccc} \hline
$n_p=0$     & $n_p=1$ & $n_p=2$ & HOMO & LUMO & Other & Rest \\
\hline 0.37 & 0.16    & 0.03    & 0.03 & 0.10 & 0.05  & 0.26 \\
\hline
\end{tabular}
\end{table}

\begin{figure}
\includegraphics[width=58.6mm]{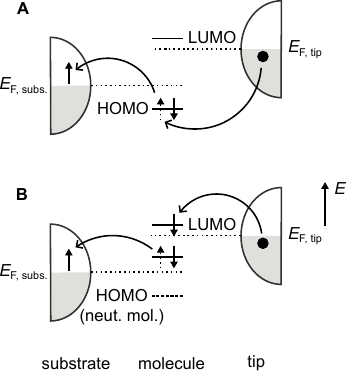}
\caption{(A) Tunneling process of the type discussed in this study.
(B) Tunneling process leading to the creation of a triplet exciton, without invoking spin-orbit coupling to flip a spin.}
\label{fig:7}
\end{figure}

This study reveals important facts for imaging of molecules and beyond.
It shows that there is access to energetically deep MOs that are inaccessible by conventional STM because voltages of several eV between tip and sample may damage the molecule or the buffer layer.
Moreover, theoretical models that exclude orbitals at energies far from the transport gap are unlikely to properly reproduce in-gap images.
While it is a widespread assumption that the electron propagates through a molecule on an exponentially decaying tunneling trajectory, analogous to how an electron tunnels through the vacuum barrier between molecule and STM tip, the above analysis shows that propagation is via electronic states which are not exponentially decaying within the extension of the molecule.
These results are particularly relevant for energy up-conversion light emission processes, like those of single isolated molecules studied with STM \cite{chenSpinTripletMediatedUpConversion2019a, farrukhBiaspolarityDependent2021, grewalSingleMolecule2022}.
Figure \ref{fig:7}A shows schematically the tunneling process discussed here for the case when the tip Fermi energy is in the transport gap.
Figure \ref{fig:7}B shows a similar process, which is the first step in an energy up-conversion process necessary to create a singlet exciton via creation of a lower energy triplet exciton.
A spin down tip electron hops into the LUMO, different from a spin up electron hopping into the HOMO in Figure \ref{fig:7}A.
It illustrates how one can create a triplet exciton state without flipping a spin, which would otherwise require invoking the very weak spin-orbit coupling or some other weak mechanism.

\section*{Conclusion}
Molecules adsorbed on thin insulating layers are supposed to behave as quasi-isolated quantum systems whose electronic structure can be studied by a scanning tunneling probe.
Here we showed clear deviations from this simple picture by analyzing the electronic states in the energy gap between HOMO and LUMO and within the transport gap of the decoupling insulator.
At these energies there exist no states of a perfectly isolated molecule, nor for an infinitely extended insulator.
The proximity of molecule, insulator, and substrate result in a continuum of real electronic states within this gap that penetrate through insulator and molecule.
Each of these states can be represented by a sum of many electronic eigenstates of the perfectly isolated system with significant weight on states even at energies far below the gap region.
We have studied PtPc and MgPc (theoretically and experimentally) adsorbed on a NaCl film on an Au(111) substrate, focusing on the states in the transport gap of these molecules.
Although PtPc and MgPc are only one atomic layer thick, the images are quite different from the image of the NaCl substrate.
Replacing PtPc with MgPc primarily rotates the image by $53^\circ$, corresponding directly to the rotation of the MgPc molecule.
This shows that the image is mainly determined by the electronic structure of the adsorbed molecule, even when the tunneling is through the gap.
We showed how the molecule's presence affects the tunneling current for three models of increasing complexity.
It is then not surprising that the electronic states of the molecule strongly influences the shape of the image.
We showed that the image is mainly determined by linear combinations of the bound states of the molecule.
We find that for energies in the gap, not too close to the HOMO or LUMO, most of the contributions come from PtPc states at energies well below the HOMO, particularly from states with no or one angular node.
Generally speaking, the NaCl film is often considered a buffer that allows access to the specific electronic \cite{reppSnellLaw2004} and topographic \cite{sunNaClMultilayer2008} properties of the substrate but ensures a sufficient electronic decoupling of an adsorbed molecule from the substrate.
We find, however, that electronic states of an electrically insulating buffer influence the image of a molecule in its transport gap substantially.
The character of the gap states is essential for more complex processes, for example, the emission of photons by a tunneling electron, where transport through the gap can play an important role.
If we treat our molecular system in essence as a generic molecule adsorbed on an insulator on a metallic substrate, we arrive at the conclusion that we can potentially access information on energetic states that are nominally inaccessible through direct tunneling.
This finding has very immediate and deep implications for imaging molecules on surfaces.

\section*{Acknowledgements}
The authors thank H. Imada and Y. Kim for providing the experimental data for topographical images of MgPc and \ce{H_2Pc} shown in Figures \ref{fig:5}C and D.

\section*{Methods and Experimental}
\setlength\parindent{0pt}\textbf{Sample preparation} -- The experiments were carried out with a home-built low-temperature STM operated at $T = 4.3$ K in an ultra-high vacuum $(<10^{-11}$ mbar) \cite{kuhnkeVersatileOptical2010}.
The Au(111) single-crystal ($>99.999$\% purity) sample was cleaned by repeated cycles of Ar$^+$ ion sputtering at $10^{-6}$ mbar range argon pressure with 600 eV acceleration energy and subsequent annealing to 873 K.
The sample heating and cooling rate was about 1 \ce{K/s}.
NaCl was evaporated thermally from a Knudsen cell held at 900 K, with the Au(111) surface held at 300 K, to obtain defect-free, (100)-terminated NaCl islands.
Next, PtPc was evaporated atop a liquid nitrogen cooled Au(111) substrate, partially covered with NaCl.
The PtPc Knudsen cell was held at 710 K while the temperature of the Au(111) substrate was about 90 K.
The sample was then transferred to the STM for characterization.
An electrochemically etched gold wire \cite{yangFabricationSharp2018} (99.95\% purity) was used as a tip in the experiment.\\
\textbf{STM measurements} -- To ensure a metallic tip, the Au wire was further prepared by controlled tip indentations ($\Delta z = 1-3$ nm, $V=50-100$ mV) in Au(111) until atomic resolution is obtained at the tunneling current set point: $I_T=10$ pA, $+1$ V.
This study always specifies bias voltages of the metal substrate with respect to the grounded tip.

\end{document}


\title{Supplementary Material for \\Character of electronic states in the transport gap of molecules on surfaces}

\maketitle

\section*{Simple model with one molecular level}
To improve the understanding of gap states, we first consider a very simple tight-binding model, where the molecule has just one (HOMO) level, and the substrate and the molecule are described in the Anderson impurity model \cite{andersonLocalizedMagnetic1961}
\begin{eqnarray}\label{eq:0a}
&&H^{\rm SH}_{0}=\sum_{\sigma} \varepsilon^{\rm H}n_{\sigma}+\sum_{{\bf k}\sigma}\varepsilon^{\rm S}_{\bf k}n^{\rm S}_{{\bf k}\sigma}  \\
&&+\sum_{{\bf k}\sigma}V^{\rm SH}\left[ c^{\dagger}_{\sigma}(c^{\rm S})^{\phantom \dagger}_{{\bf k}\sigma}+(c^{\rm S})^{\dagger}_{{\bf k}\sigma}c^{\phantom \dagger}_{\sigma}\right]. \nonumber
\end{eqnarray}
Here, the first term describes the molecular HOMO level with the energy  $\varepsilon^{\rm H}$  and occupation number $n_{\sigma}$ for electrons with spin $\sigma$.
The second term describes the substrate levels with quantum numbers ${\bf k}$ and energies $\varepsilon^{\rm S}_{\bf k}$.
The third term describes the hopping between the substrate and the HOMO level with the hopping integral $V^{\rm SH}$.
The corresponding annihilation operators are $(c^{\rm S})_{{\bf k}\sigma}$ and $c_{\sigma}$.
The substrate has a semi-elliptic density of states (DOS), $\rho(\varepsilon)$, with the width $2B$
\begin{equation}\label{eq:0b}
\rho(\varepsilon)=\sum_{\bf k}\delta(\varepsilon^{\rm S}_{\bf k}-\varepsilon)=
\frac{2}{\pi B^2}\sqrt{B^2-\varepsilon^2}\Theta(B^2-\varepsilon^2),
\end{equation}
where $\Theta(x)=1$ for $x>0$ and zero otherwise.
The tip is described by the Hamiltonian
\begin{equation}\label{eq:0c}
H^{\rm T}_{0}=\sum_{{\bf k}\sigma} \varepsilon^{\rm T}_{\bf k}n^{\rm T}_{{\bf k}\sigma},
\end{equation}
where $\varepsilon^{\rm T}_{\bf k}$ describes the tip energies.
The tip also has a semi-elliptic DOS, but it is displaced in energy by the bias, $U_{\rm bias} (-B\le U_{\rm bias}\le 0)$.
We treat the coupling between the molecule and the tip as a perturbation
\begin{equation}\label{eq:0e}
V(t)=e^{\delta t}V_0\sum_{{\bf k}\sigma}\left[(c^{\rm T})^{\dagger}_{{\bf k}\sigma}c^{\phantom \dagger}_{\sigma}
+c^{\dagger}_{\sigma}(c^{\rm T})^{\phantom \dagger}_{{\bf k}\sigma}\right],
\end{equation}
where $\left(c^{\rm T}\right)_{{\bf k}\sigma}$ annihilates an electron on the tip.
The perturbation $V(t)$ is turned on adiabatically at a time $t=t_0 \to-\infty$ with the time dependence exp$(\delta t)$, where $\delta>0$ and $\delta \to 0$ so that $|t_0|\delta \gg 1$.
The full Hamiltonian is
\begin{equation}\label{eq:1f}
H=H^{\rm SH}+H^{\rm tip}+V(t).
\end{equation}

\begin{figure*}
\includegraphics{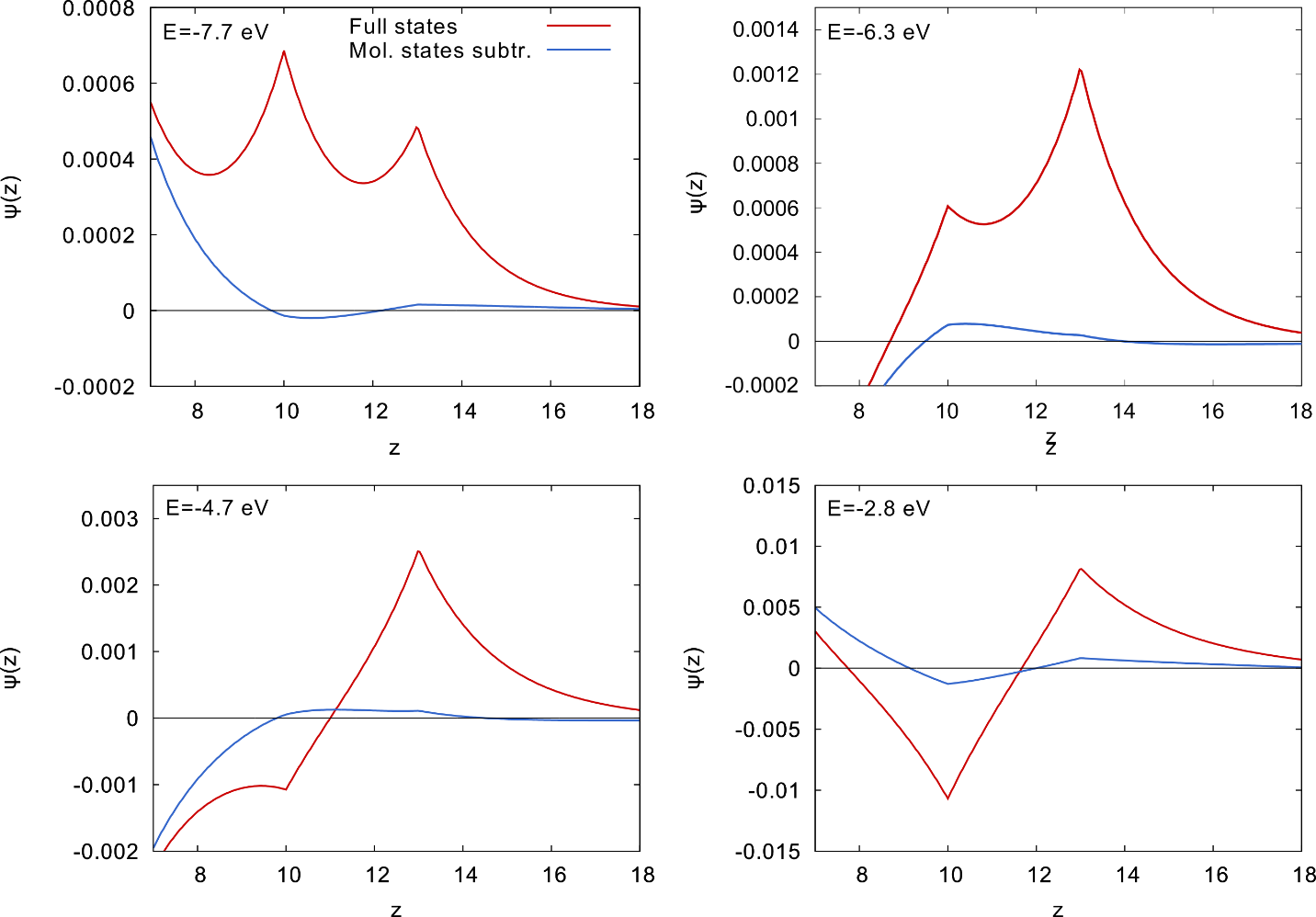}
\caption{Solutions of the $1d$ model for different energies in the range of the molecule.
We also show the rest after the free molecule solutions have been subtracted (see Eq. \ref{eq:s6}].
The rest is very small in the range of the molecule $(10 \le z \le 13)$, meaning that the two solutions provide an efficient basis set in this range and for these energies.
As the energy is increased, the solution changes from bonding to antibonding character.
Below the HOMO and above the LUMO the solution has a larger weight on the innermost atom while between these energies it has most weight on the outermost atom.
This can easily be understood by studying the interference between the terms in an expansion in the HOMO and LUMO.}
\label{fig:1}
\end{figure*}

We first solve the Anderson impurity model.
The local DOS, $\rho^{\rm HOMO}(\varepsilon)$,  on the HOMO level is
\cite{andersonLocalizedMagnetic1961}
\begin{equation}\label{eq:0f}
\rho^{\rm HOMO}(\varepsilon)=\frac{1}{\pi}{\rm Im}\frac{1}{\varepsilon-\varepsilon^{\rm H}-
\Lambda(\varepsilon)-i\Delta(\varepsilon)}
\end{equation}
where
\begin{equation}\label{eq:0g}
\Delta(\varepsilon)=\frac{2\left(V^{\rm SH}\right)^2}{B^2}\sqrt{B^2-\varepsilon^2}\Theta(B^2-\varepsilon^2)
\end{equation}
and
\begin{equation}\label{eq:0h}
\Lambda(\varepsilon)=\dfrac{2\left(V^{\rm SH}\right)^2}{\pi B^2}
\begin{cases}
\varepsilon, & \text{if}~|\varepsilon|\le B, \\
\varepsilon +\sqrt{\varepsilon^2-B^2}, & \text{if}~\varepsilon<-B, \\
\varepsilon -\sqrt{\varepsilon^2-B^2}, & \text{if}~\varepsilon>B. \\    
\end{cases}
\end{equation}
We rewrite the Anderson Hamiltonian, $H^{\rm SH}_0$, in terms of the eigenstates as
\begin{equation}\label{eq:0i}
H_0^{\rm SH}=\sum_{i\sigma} \varepsilon_i^{\rm SH}n^{\rm SH}_{i\sigma},
\end{equation}
where $\varepsilon^{\rm SH}_i$ are the eigenvalues of $H^{\rm SH}_0$ and $n^{\rm SH}_{i\sigma}$ the corresponding number operators.
We now introduce the hopping between the Anderson impurity and the tip in term of the new operators $c^{\rm SH}$
\begin{equation}\label{eq:0j}
V(t)=e^{\delta t}\sum_{ij\sigma}W_{i}\left[
\left(c^{\rm SH}\right)^{\dagger}_{i\sigma} \left(c^{\rm T}\right)^{\phantom \dagger}_{j\sigma}+
	\left(c^{\rm T}\right)^{\dagger}_{j\sigma}\left(c^{\rm SH}\right)^{\phantom \dagger}_{i\sigma}\right],
\end{equation}
where $W_{i}$ is expressed in terms of $V_0$ and the solutions $|i\rangle$ of the Anderson model
\begin{equation}\label{eq:0k}
|i\rangle=\left(a^ic^{\dagger}+\sum_{\bf k} a^i_{\bf k}c_{\bf k}^{\dagger}\right)|{\rm vacuum}\rangle.
\end{equation}
Then
\begin{equation}\label{eq:0l}
W_i=a^iV_0.
\end{equation}
We then also have that
\begin{equation}\label{eq:0m}
\rho^{\rm HOMO}\left(\varepsilon\right)=\sum_i \left|a^i\right|^2 \delta\left(\varepsilon-\varepsilon_i^{\rm SH}\right)
\equiv \left|a\left(\varepsilon\right)\right|^2\rho\left(\varepsilon\right),
\end{equation}
since $a_i$ only depends on $\varepsilon_i^{\rm SH}$.

We consider an initial state $|a\rangle$, where both the Anderson model and the tip are filled up to their Fermi levels at $\varepsilon_F^{\rm S}=0$ and $\varepsilon_F^{\rm T}= U_{\rm bias}$, respectively.
The corresponding energy, $E_a=0$, we use as energy zero. In a final state, there is a hole in the Anderson model at energy $E_i\le 0$ and an electron on the tip with energy $E_j\ge U_{\rm bias}$.
The first order perturbation theory transition amplitude between the initial state $|a\rangle$ and a state $|ij\rangle$ is given by
\begin{eqnarray}\label{eq:1}
&&\left\langle ij\left|U^{(1)}\right|a\right\rangle=\frac{1}{i\hbar}\int_{\infty}^t \mathrm{d}\tau~
e^{\left[-i\left(E_j-E_i\right)\left(t-\tau\right)/\hbar+\delta \tau\right]}W_{i}\nonumber \\
&&=\frac{e^{\delta t}}{E_j-E_i+i\delta \hbar}W_i.
\end{eqnarray}
We then have the probability
\begin{eqnarray}\label{eq:2}
&&\left|\left\langle ij\left|U^{(1)}\right|a\right\rangle\right|^2=\left|W_{i}\right|^2
\left|\frac{1}{E_j-E_i+i\delta \hbar}\right|^2e^{2\delta t}\\
&& \to \frac{\pi}{\delta\hbar}\left|W_{i}\right|^2e^{2\delta t}  \delta\left(E_i-E_j\right).
\end{eqnarray}
We then perform the sums over $i$, $j$ and spin $\sigma$
\begin{eqnarray}\label{eq:5}
&&\sum_{ij\sigma}\left|\left\langle ij\left|U^{(1)}\right|a\right\rangle\right|^2=\frac{2\pi}{\delta \hbar}\sum_i\left|W_i\right|^2\rho(E_i-U_{\rm bias})e^{2\delta t} \nonumber \\
&&=\frac{2\pi V_0^2}{\delta \hbar}\sum_i\left|a_i\right|^2\rho\left(E_i-U_{\rm bias}\right)e^{2\delta t} \\
&&=\frac{2\pi V_0^2}{\delta \hbar}\int_{U_{\rm bias}}^0 {\rm d} \varepsilon
\rho^{\rm HOMO}(\varepsilon)\rho(\varepsilon-U_{\rm bias})e^{2\delta t} \nonumber
\end{eqnarray}
The current is obtained by taking the derivative with respect to $t$.
For $t=0$ we obtain
\begin{equation}\label{eq:6}
\frac{4\pi V_0^2}{\hbar}\int_{U_{\rm bias}}^0 {\rm d}\varepsilon~\rho^{\rm HOMO}\left(\varepsilon\right)\rho\left(\varepsilon-U_{\rm bias}\right).
\end{equation}

The quantity is shown in Fig.~\ref{fig:1}, normalized to its value at $U_{\rm bias}=-2$ V.
It represents a dimensionless rate of transfer of electrons at $t=0$.

In this discussion, we have neglected the Coulomb interaction. We could introduce a new Hamiltonian for the HOMO level with the interaction $U$
\begin{equation}\label{eq:7}
H^{\rm M}=\tilde \varepsilon^{\rm H}\sum_{\sigma}n_{\sigma}+Un_{\uparrow}n_{\downarrow}.
\end{equation}
Making the transformation $\varepsilon^{\rm H}=\tilde \varepsilon^{\rm H}+U$ we obtain
\begin{equation}\label{eq:8}
H^{\rm M}=\varepsilon^{\rm H}\sum_{\sigma}n_{\sigma}+U(1-n_{\uparrow})(1-n_{\downarrow})-U.
\end{equation}
If the HOMO is well below the Fermi energy of the substrate, the state with no HOMO electron is not very important and can be neglected.
Then the Coulomb energy is not very important, except for a renormalization of the effective HOMO energy.
However, if the HOMO is close to the substrate Fermi energy, this is not true and the Coulomb energy substantially changes the physics of the problem.
The same is true if we include a LUMO in the problem.
The Coulomb interaction then leads to excitons which would otherwise be neglected.

In our tight-binding model above, we assumed that all hopping from the substrate to the tip goes via the HOMO state and that there is no direct hopping from the substrate state to the tip states.
The $1d$ model in the following section can be solved without introducing a basis set and thereby without assumptions about which
hopping integrals can be neglected.
We then show that assumptions similar to the ones above do not change the results very much.

\begin{figure*}
\includegraphics{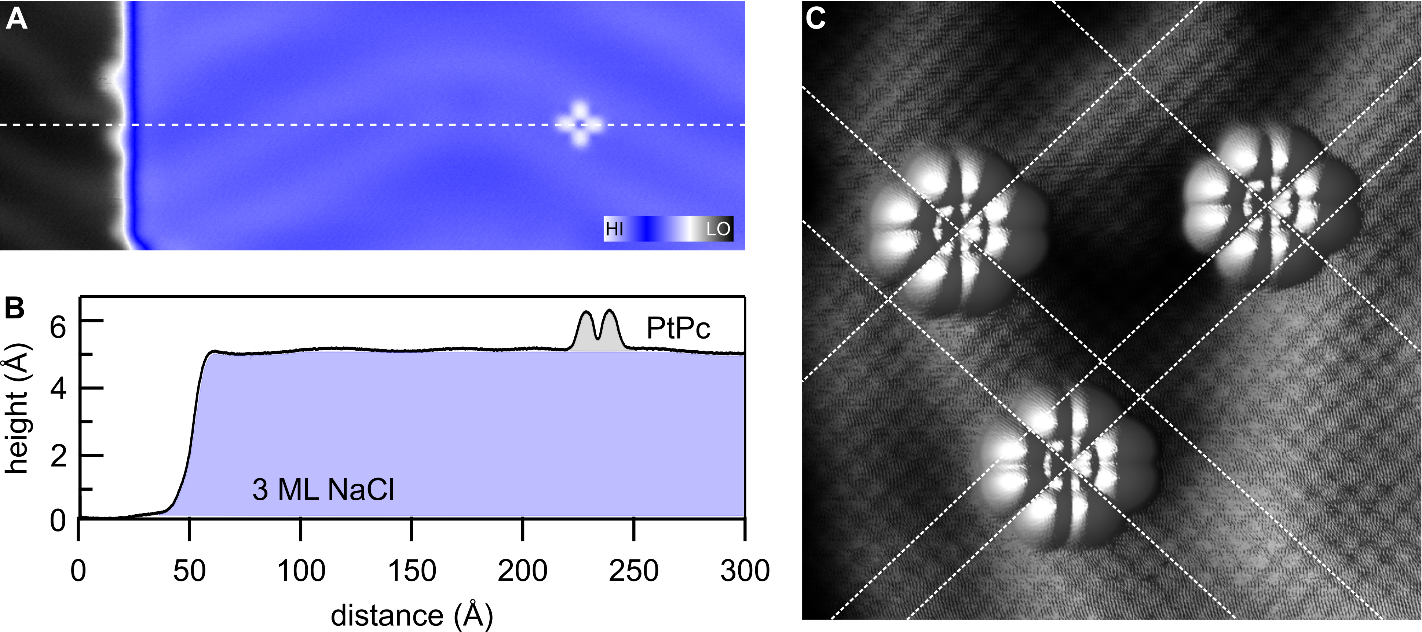}
\caption{ Overview image of STM topography for PtPc adsorbed atop 3 ML NaCl on Au(111).
(A) Overview image of PtPc atop 3 ML NaCl on Au(111) using set point: $1$ pA, $-1$ V $(300\times100 \AA^2)$.
(B) Height profile of the dashed line marked in A.
(C) STM topography image showing the HOMO of the PtPc and Cl ions of the NaCl(100) lattice using set point: $20$ pA, $-1.23$ V $(100\times100 \AA^2)$.
The white dashed lines mark the position of rows of Na ions.}
\label{fig:2}
\end{figure*}

\section*{1d model}
We now give some details of the one dimensional (1d) model in the main text.
We introduce a substrate potential
\begin{equation}\label{eq:s1}
V_{\rm substr}(z)=
\left\{\begin{array}{ll}
\infty, & {\rm for} \ z< -64 \\
-10.4, & {\rm for } \ -64\le z\le 0 \\
0, & {\rm for } \ z>0 \\
\end{array} \right.
\end{equation}
and a molecular potential
\begin{equation}\label{eq:s2}
V_{\rm mol}=V_0\left[\delta(z-z_0)+\delta(z-z0-d)\right],
\end{equation}
where $z_0=10$, $d=3$ and $V_0$ is chosen so that the levels of the free
molecule are at $-7.1$ eV and $-3.5$ eV. All lengths are measured in Bohr radii, $a_0$,
and all energies in eV. We then solve the model
\begin{equation}\label{eq:s4}
H=H_0+V_{\rm substr}+V_{\rm mol}
\end{equation}
exactly, where $H_0$ is the kinetic energy operator. These solutions are shown
in Fig.~\ref{fig:1}. We consider one energy below the HOMO, two energies
between the HOMO and the LUMO and one energy above the LUMO.

In the main text, we considered a model of PtPc on a NaCl film on Au.
These solutions are  described by a basis set, where we include Au states,
valence and conduction states of NaCl and bound states of PtPc. However, we
do  not introduce additional basis states to describe states in the gaps
of NaCl and PtPc. Since we know the exact solutions of the $1d$ model, we
can test this approximation by expanding the exact solutions in the solutions
of the free molecule in the model above
\begin{equation}\label{eq:s3}
\left(H_0+V_{\rm mol}\right)\left|i\right\rangle=E_i|i\rangle, \hskip0.3cm i=1, 2.
\end{equation}
where $|i\rangle$, $i$ = 1, 2, are the bonding and antibonding
solutions of the molecule. We then calculate
\begin{equation}\label{eq:s6}
\left|K\right\rangle -\sum_{i=1}^2\left\langle i|K\right\rangle |i\rangle,
\end{equation}
where $|K\rangle$ is a solution of the model.

In Fig.~\ref{fig:2}, we subtract the part of the solutions that can be
expanded in the HOMO and LUMO and show the reminder. The main point is that this
reminder is very small in the range of the molecule ($10\le z\le 13$) and that
the use of the HOMO and LUMO states as the only basis functions in this range is
a good approximation.

\section*{Tight-binding model}

We use a model consisting of a NaCl three-layer film on a six-layer Au substrate.
The NaCl film contains $9\times 9 \times 4 = 324$ atoms per layer.
We use three different clusters representing the Au substrate and average the results.
The three Au clusters have four, six or eight layers with 1020, 780 and 572 Au atoms per layer, respectively.
In total there are then 648 or 972 atoms in the NaCl film and 4080, 4680 or 4576 atoms in the Au substrate.
We impose periodic boundary conditions parallel to the surface for the NaCl slab and for the Au slabs.
All hopping integrals are constructed according to the rules of Harrison \cite{harrisonElementaryElectronic1999,harrison1980electronic}, including $s-d^3$ and $p-d$ hopping.
The NaCl(100) and Au(111) surfaces are non-commensurate.
We place the central Na atom on top of the central Au atom in the neighboring NaCl and Au layers.

To describe the Au substrate, we use the lattice parameter $a_{\rm Au}=4.07$~\AA~\cite{daveyPrecisionMeasurements1925}.
We use the Harrison level energies $\varepsilon_{6s}=-6.98$ eV and $\varepsilon_{5d}=-17.78$ eV as a starting point and add a $6p$ level at 5 eV above the $4s$-level.
We then shift the $5d$-level relative to the $6s$ and $6p$ so that the top of the $5d$ band is placed at 1.7 eV below the Fermi energy \cite{sheverdyaevaEnergymomentumMapping2016a}.
Finally all energies are shifted so that the Fermi energy is at zero.
These parameters are summarized in Table~\ref{table:1}.

To describe the NaCl film, we follow our earlier work \cite{leonAnionicCharacterConduction2022} and chose parameters such that the conduction band has mainly Cl $4s$ character \cite{deboerOriginConduction1999a, leonAnionicCharacterConduction2022}.
For this purpose, we replace the Cl $3s$ level by a $4s$ level, which has been strongly lowered by the Madelung potential, while the Na levels are shifted strongly upwards.
We adjust the Harrison parameters so that the experimental band gap (8.5 eV \cite{pooleElectronicBand1975b}) is reproduced for bulk NaCl.
According to the calculations \cite{wangQuantumDots2017a} using the GW method \cite{hedinNewMethod1965b}, the top of the valence band is 5 eV below the Fermi energy.
We then shift all the NaCl energies relative to the Au energies correspondingly.
The resulting parameters are summarized in Table \ref{table:1}.
The calculations were performed using the lattice parameter $a_{\rm NaCl}=5.54$ \AA \ \cite{chenPropertiesTwodimensional2014b}.

We use the calculated separation $d_{\rm Au - NaCl} = 3.12$~\AA between the Au surface and the NaCl film \cite{chenPropertiesTwodimensional2014b}.
Since the NaCl film and the substrate are incommensurate, several Au atoms can have similar distances to a given NaCl atom, and the nearest neighbors are poorly defined.
We then use a smooth distance dependent cut off of the Harrison prescription for the hopping between the substrate and the film.
Thus the Harrison prescription for these hopping integrals is multiplied by a factor
\begin{equation}
    \exp{\left(-\frac{\left(d-d_{\rm Au}-d_{\rm NaCl}\right)^2}{\lambda^2_{\rm SB}}\right)},
\end{equation}
where $d$ is the distance between an Au atom and a NaCl atom at the Au-NaCl interface.
Here, $\lambda_{\rm SB}$ is chosen such that summing these factors over all the Au neighbors of a NaCl atom and averaged over the NaCl atoms in the innermost layer adds up to four.
Then the innermost NaCl atoms effectively couple to four Au atoms.

\begin{table}[b]
\begin{tabular}{lccc}
\hline
\hline
Element &  $s$  & $p$ & $d$ \\
\hline
Au ($6s$, $6p$, $5d$) & $4.1$ & $9.1$ & $-3.7$ \\
Na ($3s$, $3p$) & $12.8$ & $16.8$ & $-$ \\
Cl ($4s$, $3p$) & $10.2$ & $-5.0$ & $-$ \\
C ($2s$, $2p$) & $-19.38$ & $-11.07$ & $-$ \\
N ($2s$, $2p$) & $-26.22$ & $13.84$ & $-$ \\
Pt ($6s$, $5d$) & $-6.85$ & $-$ & $-16.47$  \\
Mg ($3s$, $3p$) & $-6.89$ & $-3.79$ & $-$ \\
H ($1s$) & $-13.61$ & $-$ & $-$ \\
\hline
\hline
\end{tabular}
\caption{Level energies used for PtPc or MgPc absorbed on a NaCl film on a Au substrate.}
\label{table:1}
\end{table}

We study the absorbed molecules platinum phthalocyanine (PtPc) and magnesium phthalocyanine (MgPc).
The coordinates of PtPc are  obtained from a density functional calculation.
The same coordinates are used for MgPc.
The tight-binding parameters are obtained from Harrison \cite{harrisonElementaryElectronic1999} and are given in Table~\ref{table:1}. For the H atoms, we include the $1s$ level at the energy $-13.6$ eV (not given by Harrison).
Guided by Miwa {\it et al.} we use the separation 3.4 \AA \ between the the molecules and the NaCl film both for PtPc and MgPc.
MgPc is absorbed on top of a Cl atom \cite{miwaEffectsMoleculeinsulator2016a}, while PtPc on top of a Na atom (see Fig.~\ref{fig:2}D).
For PtPc, the four "arms" of the molecule are along the NaCl (100) directions, while for MgPc they are close to the NaCl(110) directions with a small deviation of 8$^{\circ}$ included in the calculation \cite{miwaEffectsMoleculeinsulator2016a}.
Again a $\lambda_{\rm BM}$ is chosen so that, on average, from each atom in the molecule there is effectively hopping to four atoms in the NaCl buffer.
The Au slab breaks the four-fold symmetry of PtPc and MgPc which has been reintroduced in the plots.

For the PtPc molecule, these parameters incorrectly put a $\sigma$-orbital below the HOMO.
We, therefore, shift this orbital upwards by 3.2 eV.
The parameters in Table \ref{table:1} lead to too small a gap.
We then shift the occupied levels downwards so that the HOMO of the free molecule is at $-1.42$ eV and the unoccupied levels upward so that the LUMO is 1.69 eV.
After the interactions with the substrate and NaCl are included, the HOMO is then at $-1.3$ eV and the LUMO at $1.7$ eV ($E_{\rm F}=0$), in agreement with experiment.

\section*{Experimental details}
Figure~\ref{fig:2}A shows an overview image of the area where experimental results shown in Fig. 5 in the main text are obtained.
The topographic height of NaCl as determined from fitting an error function to the line profile is $4.74\pm0.01$~\AA~(shown in Fig.~\ref{fig:2}B).
The in-gap image of PtPc atop 3 ML NaCl shows a sample voltage dependence and has a topographic height of $\approx 0.6-1.2$~\AA~for sample voltage of $-0.2$ to $-1$ V.
Figure \ref{fig:2}C shows the STM topography image of HOMOs of the PtPc atop 3 ML NaCl atop Au(111).
The higher set point used (20 pA), allows simultaneous imaging of the Cl ions of NaCl(100) lattice.
The white dashed lines in Fig.~\ref{fig:2}D marks the Na ion rows of the NaCl(100).
The intersection of two Na ion rows overlapping with center of the PtPc HOMO topography indicates that PtPc is adsorbed atop the Na ion.
We emphasize that the experimental results for PtPc are obtained using a metallic Au tip which is distinct from the the tip with a carbon monoxide (CO) tip employed by Miwa {\it et al} \cite{miwaEffectsMoleculeinsulator2016a}.

\begin{figure}
\includegraphics{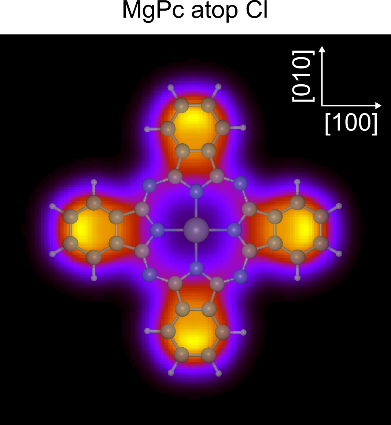}
\caption{Theoretical result for MgPc adsorbed atop Cl as in Ref.~\onlinecite{miwaEffectsMoleculeinsulator2016a} but orientated along (010) axis of the NaCl. The orientation is similar to that of PtPc or H$_2$Pc. Size: $20\times20 \AA^2$}
\label{fig:3}
\end{figure}

\section*{Comparing in-gap maps in constant current and constant height modes}
In the main text, we compare the calculations to constant height scans in Figure 4 and to constant current scans in Figure 5.
In the following we show that there is no qualitative difference between in-gap images in constant height and constant current mode.
Note that in the vacuum tail of the electronic wave function probed by STM, the current depends exponentially on the distance, which provides the basis for the functional relation between the different types of maps.
Plotting both maps with the same color scale overemphasizes the contrast in constant height maps and lets weak features often disappear within experimental noise.
In Figure 4 below we compare calculations (A, B) with constant height maps (C, D) and constant current maps (E, F).
We find that by using the same linear color scales, the constant height maps show a closer resemblance to theory because the local density of states in the calculation is also evaluated at a constant height above the molecule.
However, the observe ``cross'' shape of the in-gap feature is the same in all cases.

\begin{figure}
\includegraphics[width=\linewidth]{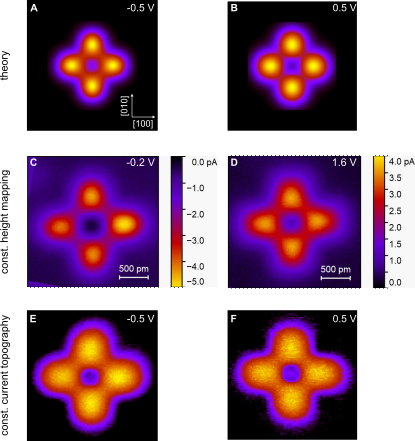}
\caption{Maps of PtPc molecules in the transport gap for sample voltages given in the top right corner of each panel.
(A,B) Theoretical results.
(C,D) Experimental constant height maps of a molecule with current scales given on the right of each panel.
(E,F) Experimental constant current maps on a different molecule.
The length scale for all panels is given in (C) and (D).}
\label{fig:4}
\end{figure}

\section*{Scanning tunneling spectroscopy in the transport gap}
In Figure 5 below, we compare d$I$/d$V$ spectra obtained by theory (A) with experimental data (B, C). 
The dynamics of the STM data is limited by instrumental noise typically to about 1 pA/V in spectra that allow to capture in-gap features together with the signal of the frontier orbitals.
In Figure 5B and 5C, we use the numerical derivative of the current signal in order to avoid the zero offset, which is typically present in data from a lock-in amplifier.
By doing this, we can identify in Figure 5B a value of 2 pA/V around zero sample voltage which can be recognized as a true signal indicated by the reduced scattering of data points on the logarithmic scale.
In Figure 5C, this signal is even more pronounced.
The in-gap d$I$/d$V$ signal is about 2-3 orders of magnitude below the signal at the top of the HOMO and thus close to the result of the calculation.
The pronounced step in Figure 5C near 0 eV may be due to the additional electronic density of states from the Au-NaCl interface state, which has a sharp onset around $-0.25$ eV \cite{lauwaetDependenceof2012}.
At positive voltages, the calculation exhibits the step around $+1$ eV, which we attribute to the same interface state which is, however, shifted due to the employed vacuum level alignment of energies which may not agree with the experimental alignment between substrate and molecular states.

\begin{figure}[b]
\includegraphics[width=\linewidth]{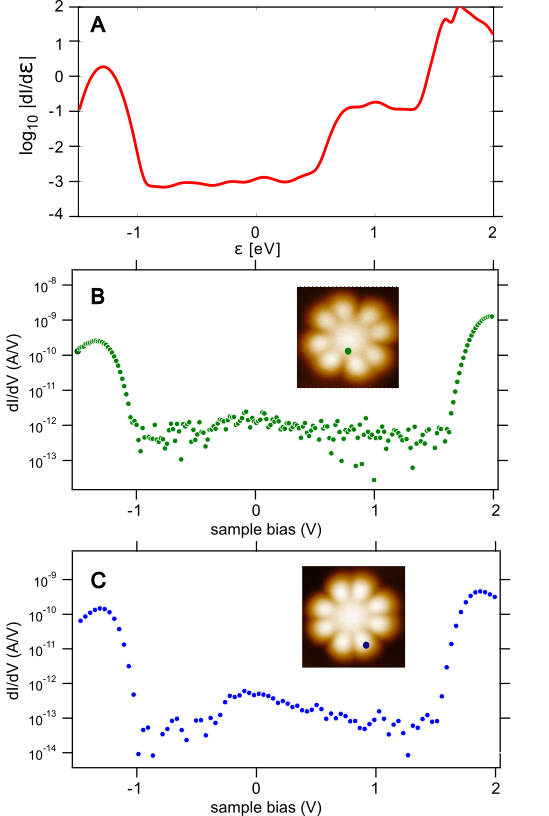}
\caption{Differential conductance from theory for a PtPc molecule on 3 ML of NaCl averaged over the entire molecule. 
(B, C) STM current signal numerically differentiated with respect to bias voltage.
The spectra are local measurements on the positions marked by the dot on the respective inset.}
\label{fig:5}
\end{figure}